\documentclass[showpacs,aps,prl,twocolumn,superscriptaddress]{revtex4-1}
\usepackage{graphicx} 
\usepackage{dcolumn}
\usepackage{bm}
\usepackage{array}
\usepackage{amssymb,amsmath}
\usepackage{epstopdf}
\usepackage{color}
\usepackage{mathrsfs}
\usepackage[normalem]{ulem}

\newcolumntype{C}[1]{>{\centering\arraybackslash}p{#1}}
\definecolor{mygrey}{gray}{0.75}

\begin{document}
\title{Revealing temperature evolution of the Dirac band in ZrTe$_5$ via magneto-infrared spectroscopy}

\author{Yuxuan Jiang}
\email{yuxuan.jiang@ahu.edu.cn}
\affiliation{School of Physics and Optoelectronics Engineering, Anhui University, Hefei 230601, China}
\affiliation{Center of Free Electron Laser and High Magnetic Field, Anhui University, Hefei 230601, China}
\author{Tianhao Zhao}
\affiliation{School of Physics, Georgia Institute of Technology, Atlanta, Georgia 30332, USA}
\author{Luojia Zhang}
\affiliation{School of Physics, Georgia Institute of Technology, Atlanta, Georgia 30332, USA}
\author{Qiang Chen}
\affiliation{Department of Physics and Astronomy, University of Tennessee, Knoxville, Tennessee 37996, USA}
\author{Haidong Zhou}
\affiliation{Department of Physics and Astronomy, University of Tennessee, Knoxville, Tennessee 37996, USA}
\author{Mykhaylo Ozerov}
\affiliation{National High Magnetic Field Laboratory, Tallahassee, Florida 32310, USA}
\author{Dmitry Smirnov}
\affiliation{National High Magnetic Field Laboratory, Tallahassee, Florida 32310, USA}
\author{Zhigang Jiang}
\email{zhigang.jiang@physics.gatech.edu}
\affiliation{School of Physics, Georgia Institute of Technology, Atlanta, Georgia 30332, USA}
\date{\today}

\begin{abstract}
We report the temperature evolution of the Dirac band in semiconducting zirconium pentatelluride (ZrTe$_5$) using magneto-infrared spectroscopy. We find that the band gap is temperature independent at low temperatures and increases with temperature at elevated temperatures. Although such an observation seems to support a weak topological insulator phase at all temperatures and defy the previously reported topological phase transition (TPT) at an intermediate temperature in ZrTe$_5$, we show that it is also possible to explain the observation by considering the effect of conduction-valence band mixing and band inversion with a strong topological insulator phase at low temperatures. Our work provides an alternative picture of the band gap evolution across TPT.
\end{abstract}

\maketitle
\begin{figure}[t!]
\centering
\includegraphics[width=3.25in] {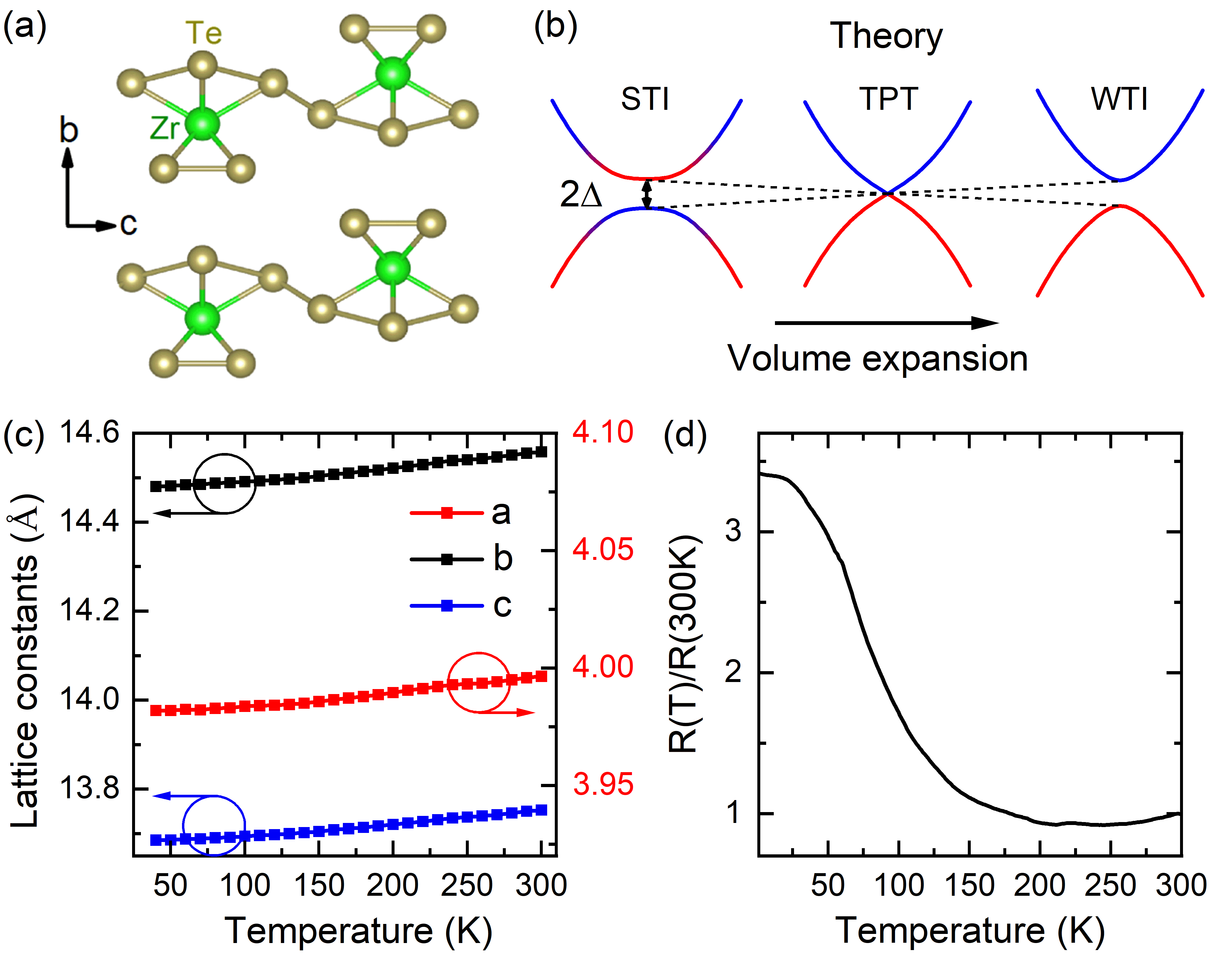}
\caption{(a) Crystal structure of ZrTe$_5$ in the $b$-$c$ plane. (b) Schematic illustration of a volume expansion induced TPT and the associated low-energy band structure around $\Gamma$ point. Here, $2\Delta$ is the band gap. The WTI phase is expected to have a normal band gap in the bulk, similar to that of NI. The red and blue colors denote the conduction and valence band characters in different energy bands. (c) Temperature dependence of the lattice constants along different crystallographic directions, determined by x-ray powder diffraction measurements. (d) Temperature-dependent resistance for the Te-flux grown ZrTe$_5$. The curve is normalized to the resistance value at 300 K.}
\label{fig1}
\end{figure}

ZrTe$_5$ is a van der Waals material with the layer stacking direction along the $b$-axis of the crystal (Figure \ref{fig1}a). The recent interest in ZrTe$_5$ originates from the theoretical prediction of a room-temperature quantum spin Hall insulator phase in its monolayer limit and a three-dimensional topological insulator (TI) phase in its bulk form \cite{Weng2014transition}. The prediction has sparked intensive experimental investigations into both the electronic and topological properties of ZrTe$_5$ \cite{Li2016Chiral,Liang2018anomalous,liu2021induced,gourgout2022magnetic,lozano2022anomalous,Tang2019three,galeski2021origin,wei2022extremely,tian2021gap,ehmcke2021propagation,Chen2015optical,Wu2016evidence,Li2016experimental,Manzoni2016evidence,Chen2017spectroscopic,Jiang2017landau,Tian2017Weyl,zhang2021observation,zhu2022comprehensive}. However, different topological phases, \textit{i.e.}, weak/strong TIs (WTI/STI) and Dirac/Weyl semimetals, have all been reported in ZrTe$_5$ from different experiments \cite{Li2016Chiral,Liang2018anomalous,Chen2015optical,Wu2016evidence,Li2016experimental,Manzoni2016evidence,Chen2017spectroscopic,Jiang2017landau,Tian2017Weyl,zhang2021observation,zhu2022comprehensive}. Such discrepancy may result from the sensitive dependence of the topological phase on the lattice constants \cite{Weng2014transition,Fan2017transition}. In addition, many theoretical calculations have predicted that volume expansion can result in a topological phase transition (TPT) in ZrTe$_5$ \cite{Fan2017transition,Weng2014transition,Wang2020ultrafast,zhang2021observation}, as schematically illustrated in Figure 1b. These results invite controllable measurements across the TPT to reconcile the experimental observations.

While several different techniques exist to manipulate the lattice constants in materials (such as using strain \cite{mutch2019evidence} or ultrafast laser \cite{Wang2020ultrafast,konstantinova2020photoinduced}), temperature remains the most convenient method, and it has been used to explore the topological phases in ZrTe$_5$ \cite{Manzoni2016evidence,Zhang2017electronic,Xu2018temperature}. Unfortunately, in these experiments, the band gap behavior across the TPT remains elusive. For example, high-resolution angle-resolved photoemission spectroscopy measurements find that the band gap never closes in the temperature range of $2<T<255$ K \cite{Zhang2017electronic,manzoni2017temperature,xiong2017three,song2022temperature,zhang2021observation}, while a zero-field infrared (IR) optical conductivity measurement reveals gap closure at an intermediate temperature \cite{Xu2018temperature}. In the latter case, the band gap evolution has been further connected to an anomalous resistance peak in $R(T)$ at a critical temperature $T^*$ \cite{Okada1980anomaly}. However, such a connection is still under debate as it is not consistently reported between different experimental techniques \cite{Tian2019dirac,Zhang2017electronic}. In addition, the presence of resistivity anomaly in ZrTe$_5$ can be eliminated by using the flux growth method, as it can effectively reduce the amount of Te vacancies in the material \cite{Shahi2018bipolar}. Even though there is a growing interest in flux-grown samples for their better quality and intrinsic behavior \cite{Shahi2018bipolar,mutch2019evidence,wang2022gigantic,sun2020large,wang2018discovery}, the temperature-dependent band gap study is still lacking.
 
In this work, we examine the band structure evolution in temperature of the molten Te-flux grown ZrTe$_5$ single crystals using magneto-IR spectroscopy. As discussed earlier, the flux growth method helps eliminate the effect of resistivity anomaly. More importantly, magneto-IR spectroscopy can directly probe the band structures of different carriers with high accuracy by tracing their Landau level (LL) transitions. It helps alleviate the complications caused by Fermi level shifting and other thermally excited carriers. By tracing the Dirac-like band in ZrTe$_5$ up to 175 K, we observe a nonlinear monotonic increase in band gap with increasing temperature. Even though it is possible to explain such a behavior with a no TPT scenario, we also propose a mechanism to reconcile this observation with our prior results that are compatible with the existence of a TPT. We argue that this behavior is caused by the band inversion and the associated orbital mixing effect. Our work sheds light on the importance of the conduction-valence band mixing in describing the TPT.

The Te-flux growth of ZrTe$_5$ single crystals is described in our previous work \cite{Jiang2020unraveling}. Figure \ref{fig1}d shows the normalized temperature-dependent resistance, $R(T)/R(300\text{K})$, of our sample, where no visible anomalous resistance peak is observed except at very low temperatures as reported in Refs. \cite{Shahi2018bipolar,mutch2019evidence,wang2022gigantic,sun2020large,wang2018discovery}. X-ray powder diffraction measurements also find that the thermal expansion of the lattice is smooth in the temperature range of $40<T<300$ K (Figure \ref{fig1}c). No structure change is spotted. For magneto-IR measurements, we repeatedly exfoliate a bulk crystal over an IR-transparent Scotch tape to achieve maximum coverage of the tape. The sample/tape composite is then placed on a metal aperture with a heater wrapped around it and a temperature sensor on the backside. After loading the sample into the magnet (in Faraday geometry), the transmission spectra are taken with a Fourier transform IR spectrometer, where the transmitted IR intensity from a Globar light source is detected by a Si bolometer shortly behind the sample. The temperature range of the sample is varied from 6 K to 175 K.

\begin{figure*}[t!]
\centering
\includegraphics[width=\linewidth] {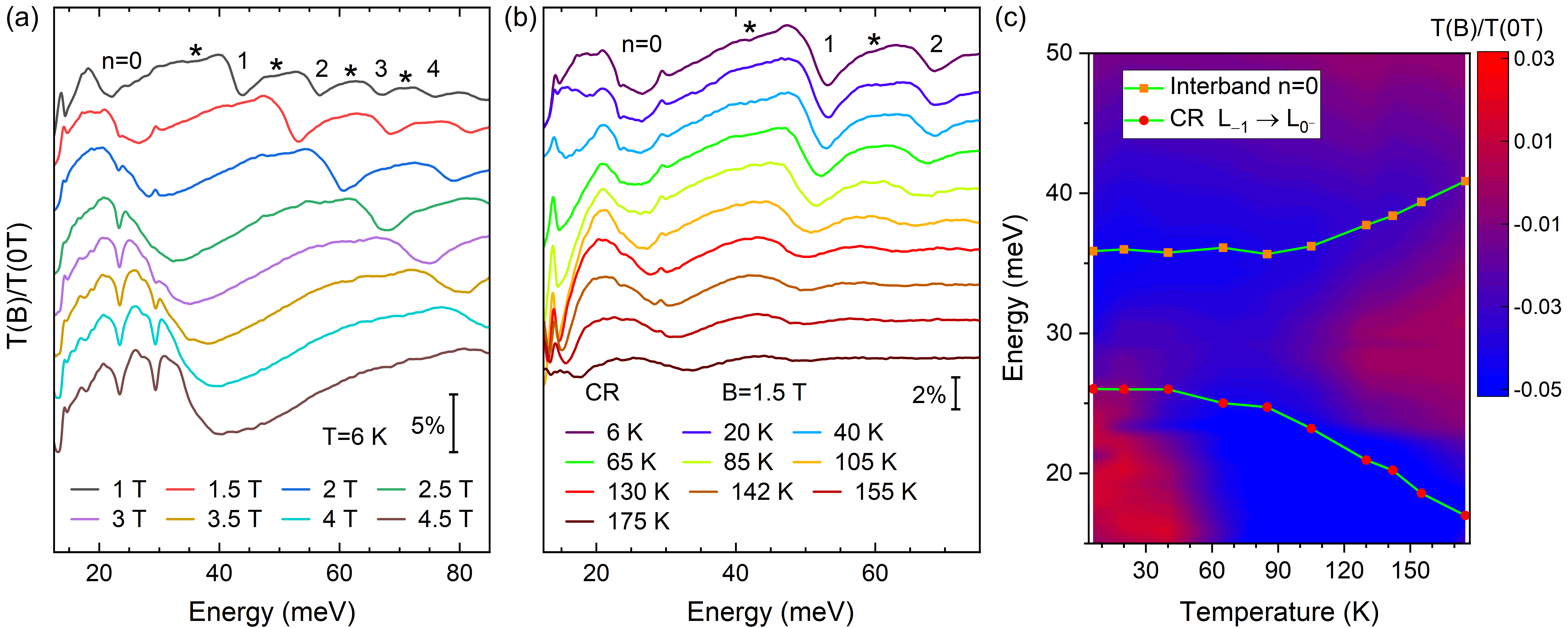}
\caption{(a) Magnetic field dependence of normalized magneto-transmission spectra, $T(B)/T(0\text{T})$, of ZrTe$_5$ measured at 6 K from 1 T to 4.5 T. (b) Temperature-dependent magneto-transmission spectra, $T(B)/T(0\text{T})$, of ZrTe$_5$ from 6 K to 175 K at 1.5 T. The integer index $n$ labels the interband LL transitions $L_{-n (-n-1)}\rightarrow L_{n+1 (n)} $. The asterisk symbol ($*$) indicates the second set of LL transitions, a hallmark of the STI phase in ZrTe$_5$ at low temperatures \cite{Jiang2020unraveling}. A CR mode emerges at elevated temperatures, which can be attributed to the $L_{-1}\rightarrow L_{0^-}$ transition. (c) False color map of the temperature evolution of the normalized magneto-transmission spectra, $T(B)/T(0\text{T})$, of ZrTe$_5$ measured at 3 T. The symbols and lines denote the CR (red symbol) and interband $n=0$ (orange symbol) transition energies calculated from the band parameters extracted from the model fitting of the interband LL transitions. The corresponding raw spectra of (c) are shown in Figure \ref{fig6} in Appendix. All measurements are performed with $B$$\parallel $$b$-axis. The spectra are offset vertically for clarity in (a) and (b).}
\label{fig2}
\end{figure*}

Figure \ref{fig2}a shows the magnetic field dependence of the normalized magneto-transmission spectra (i.e., $T(B)/T(0\text{T})$) of ZrTe$_5$ at the lowest temperature of 6 K. A series of strong dips can be observed once the magnetic field is applied, which blueshifts with increasing the field. It has been well established that these absorption dips arise from the optical transitions between the LLs in ZrTe$_5$, and one can employ the LL index $n$ to label the allowed interband transitions $L_{-n(-n-1)}\rightarrow L_{n+1(n)}$ following Refs. [\onlinecite{Jiang2017landau,Jiang2020unraveling,CHen2015magnetoinfrared,Chen2017spectroscopic,Martino2019two}]. Other prominent features in Figure \ref{fig2}a include a weaker set of LL transitions labeled by the asterisk symbol ($*$, to be discussed later) and sharp dips between 20-40 meV originating from the IR-active phonon modes \cite{Chen2015optical,Xu2018temperature}. The energies of the phonon modes do not change as a function of magnetic field, and they do not contribute to the electronic structure of ZrTe$_5$ studied in this work.

Figure \ref{fig2}b shows a typical temperature evolution of the normalized magneto-transmission spectra, $T(B)/T(0\text{T})$, at 1.5 T and from 6 K to 175 K (more temperature dependence data can be found in Figure \ref{fig5} in Appendix). Here, the transmission spectra are normalized to the zero-field spectra at each temperature to avoid background complications from the temperature change. As temperature increases, the LL transitions start to smear out or weaken due to the thermal broadening effect. At the highest temperature (175 K), only the $n \leq 1$ LL transitions remain visible. The temperature evolution of the LL transitions exhibits different behavior for different modes. With increasing temperature, the $n=0$ mode shifts toward higher energy, while the $n\geq 1$ modes shift toward lower energy within the measurement range. In addition, a new mode emerges at elevated temperatures and forms a well-defined dip at $\sim$17 meV at the highest temperature. Based on the energy of this mode, one can attribute it to the cyclotron resonance (CR) mode $L_{-1}\rightarrow L_{0^-}$, given that the sample is hole-doped in the measurement temperature range \cite{Shahi2018bipolar}. In Figure \ref{fig2}c, we present the false color map of the temperature dependence of the normalized transmission at 3 T, where the CR and $n=0$ modes are more prominent (the corresponding raw spectra are presented in Figure \ref{fig6} in Appendix). One can clearly see that these two modes separate from each other at around 60$\ $K. 

From the magneto-IR spectroscopy data (Figure \ref{fig2}), one can also extract information about the mobility ($\mu$) and carrier density in our samples. The formation of distinct LLs requires the semiclassical condition $\mu B>1$ to be satisfied, which sets the lower bound in mobility estimation \cite{Poumirol2013PRL}. For all the temperatures studied in this work, the LL transitions can be identified at as low as 1 T. Hence, $\mu>10,000$ cm$^2$V$^{-1}$s$^{-1}$ up to $T=175$ K. Furthermore, even though the emergence of the CR mode implies an increase in carrier density at elevated temperatures, the observation of the $n=0$ transition in Figure \ref{fig2} suggests that our samples enter the quantum limits at $B<1$ T for the entire temperature range and sets the upper limit for the carrier density. Therefore, we can deduce that the Dirac band in our sample remains a very low carrier density within the measurement temperature range. Our observation is in contrast to the density extracted from the reported temperature-dependent Hall measurements, which increases about two orders of magnitudes with increasing temperature \cite{Shahi2018bipolar}. To reconcile these facts, we emphasize that electronic transport measurement probes the contributions from all bands crossing the Fermi level whereas magneto-IR measurement focuses on a specific band. The discrepancy in carrier density suggests that the dominant contribution in thermally activated carriers in the Hall measurements is from other trivial bands in ZrTe$_5$, while the Dirac electrons retain its high mobility and low density up to 175 K. We note that our work is the first to show a clear LL formation at such elevated temperatures in ZrTe$_5$ using magneto-optics. As a comparison, magneto-transport measurement can only probe LLs up to around 20 K \cite{Yu2016quantum,Wang2021NC}. These results suggest that ZrTe$_5$ holds great promise in optoelectronics applications with Dirac electrons and could be further manipulated by magnetic fields. 

Next, we turn to a quantitative analysis of the band structure evolution in ZrTe$_5$ as a function of temperature. We first investigate the strong interband LL transitions labeled by integer $n$ in Figure \ref{fig2}a,b. It has been well established that at low temperatures, these transitions can be described by a simple massive Dirac fermion model
\cite{Jiang2017landau,Jiang2020unraveling,Martino2019two,CHen2015magnetoinfrared}, and the LL energies read
\begin{equation}
\label{Dirac}
E_n=\alpha \sqrt{2 e\hbar v^2_F n B+\Delta^2},
\end{equation}
with $\alpha=\pm 1$ is the band index, $e$ the electron charge, $\hbar$ the reduced Planck's constant, and $\Delta$ the Dirac mass. The allowed LL transitions satisfy the conventional selection rule $\Delta n=\pm1$, and their transition energies can be calculated accordingly. From eq. \eqref{Dirac}, one can see that the change in band gap ($2\Delta$) mostly impacts the energy of the $n=0$ LL transition, and the effect quickly becomes unnoticeable for a higher $n$ as the dominant energy scale for higher LL transitions is determined by $v_F$. For ZrTe$_5$, we have $2|\Delta|=10$ meV and  $\sqrt{2e\hbar v_F^2 B}=17.4$ meV at $B=1$ T and $T=6$ K. Therefore, we will consider the temperature dependence of both $\Delta(T)$ and $v_F(T)$ to understand the different temperature evolution observed in Figure \ref{fig2}b,c for different modes. Specifically, we reveal that the blueshift of the $n=0$ mode with increasing temperature is dominated by $\Delta(T)$, while the redshift of the $n>1$ modes is due to $v_F(T)$. The temperature dependence of the $n=1$ mode reflects the competition between the two effects. 

\begin{figure}[t]
\centering
\includegraphics[width=3.25in] {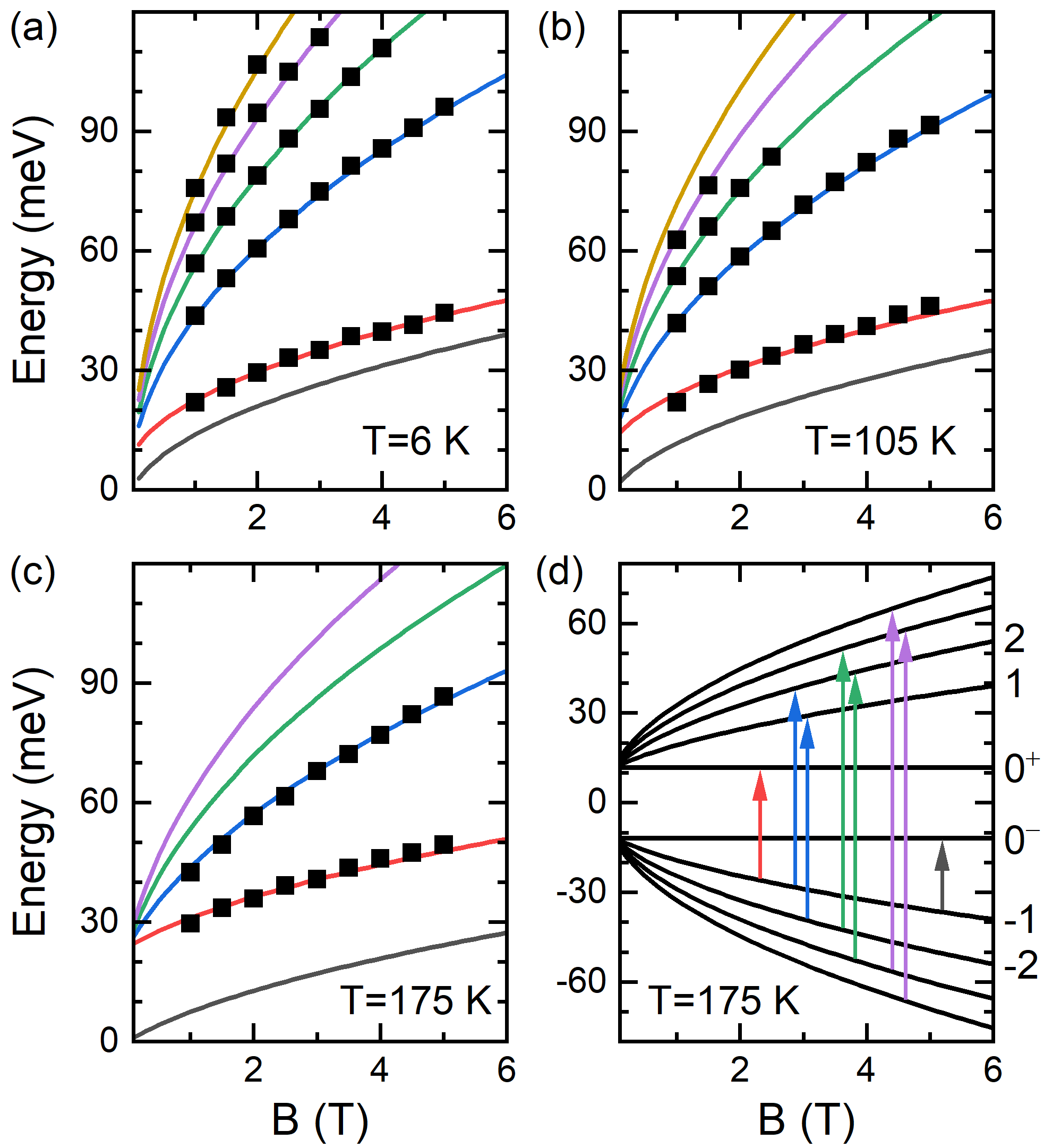}
\caption{Magnetic field dependence of LL transitions at selected temperatures of (a) $T=6$ K, (b) 105 K, and (c) 175 K. The black squares are the experimentally extracted transition energies, and the size of the squares is larger than the uncertainty of the transition energies. The solid lines are the best fits to the data using the massive Dirac fermion model \eqref{Dirac}.  (d) Representative Landau fan diagram for $T=175$ K. The LLs are labeled by the index $n=0,1,2,...$, and their corresponding LL transitions $L_{-n(-n-1)}\rightarrow L_{n+1(n)}$ are color-coded consistent with those in (a)-(c). Due to the electron-hole symmetry at low fields, the $\Delta n=\pm1$ transitions degenerate.}
\label{fig3}
\end{figure}

We then extract the energies of interband LL transitions at selected magnetic fields and plot the magnetic field dependence of each mode at specific temperatures. Figure \ref{fig3}a-c shows the extracted data (symbols) at three different temperatures as well as the fits (solid lines) using eq. \eqref{Dirac}. Excellent agreement between the experiment and model calculation is achieved throughout the measured temperature range, verifying the assignment of LL transitions. Figure \ref{fig3}d illustrates the calculated Landau fan diagram at $T=175$ K and the corresponding LL transitions following the color code in Figure \ref{fig3}a-c.

\begin{figure}[t]
\centering
\includegraphics[width=3.25in] {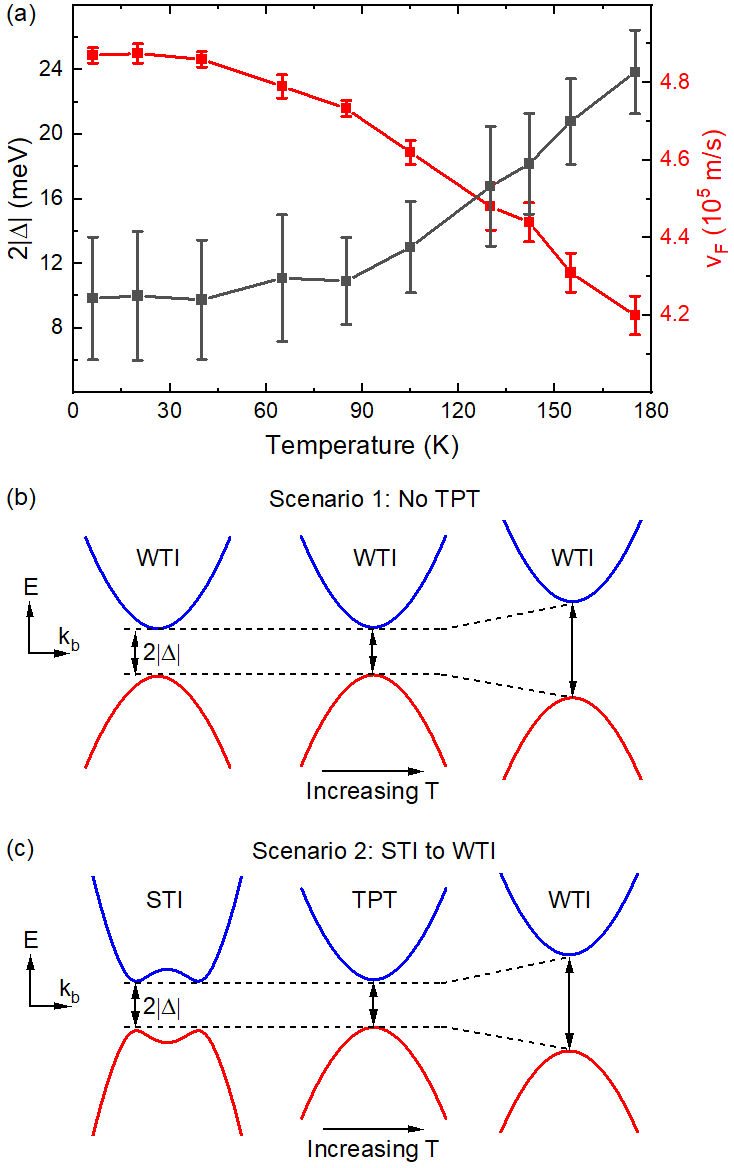}
\caption{(a) Temperature dependence of the extracted Fermi velocities and band gaps. The error bars are determined from the LL transition energy uncertainties. (b,c) Schematic drawings of the band structure evolution as a function of temperature for two possible scenarios that can describe our experiment.}
\label{fig4}
\end{figure}

Figure \ref{fig4}a summarizes the temperature dependence of the extracted band gaps and Fermi velocities. We find that $2|\Delta|$ remains almost unchanged up to around 60 K, and further increase in temperature leads to an increase in $2|\Delta|$ and an almost concurrent decrease in $v_F$. Using these parameters, we can confirm that the origin of the additional energy mode labeled by ``CR'' in Figure \ref{fig2}b is indeed the cyclotron resonance mode $L_{-1}\rightarrow L_{0^-}$. From eq. \eqref{Dirac}, the energy difference between the interband $n=0$ transition and the CR should be equal to the band gap. At low temperatures, the band gap is small, and the CR and interband $n=0$ transitions merge together. As the band gap increases at elevated temperatures, the distance between the two modes starts to increase, and therefore the CR mode is observed. We note that although it is difficult to accurately extract the CR energy due to the lineshape distortion caused by nearby phonon modes, one can instead calculate the CR and $n=0$ transition energies at $B=3$ T, using the extracted band parameters from the model fitting of the interband LL transitions. The calculation results are shown as symbols in Figure \ref{fig2}c (and also in Figure \ref{fig6} in Appendix). As one can see, the calculated energy positions correctly capture the trend of the optical weight change as a function of temperature, further confirming the origin of the CR mode and the temperature evolution of the band gap and Fermi velocity.

Having firmly established the temperature dependence of the band parameters in ZrTe$_5$, we can now discuss its implication in TPT. For the band gap, theory predicts that as temperature increases, the volume expansion may lead to two possible scenarios, depending on the topological phase at the base temperature \cite{Fan2017transition}. For scenario 1, if ZrTe$_5$ is a WTI at the base temperature, $2|\Delta|$ is expected to increase monotonically with temperature, and no TPT exists in ZrTe$_5$ (Figure \ref{fig4}b). For scenario 2, if ZrTe$_5$ is a STI at the base temperature, $2|\Delta|$ is expected to first decrease with increasing temperature to zero gap and then increase when it crosses the TPT from the STI to the WTI phase.

Based on the theoretical prediction, scenario 1 (i.e., no TPT) seems to offer a possible explanation of our experimental data as the band gap evolution in temperature shows a monotonic increase with no sign of closure. Consequently, one can assign a WTI phase to ZrTe$_5$ throughout the entire temperature range and even up to room temperature, as further increasing the temperature will continue to expand the volume. Such an interpretation is consistent with the literature reporting a WTI phase at low temperatures \cite{zhang2021observation,xiong2017three,zhu2022comprehensive}

However, we should also emphasize that there are other works reporting a STI phase in ZrTe$_5$ \cite{Manzoni2016evidence,Chen2017spectroscopic,manzoni2017temperature,Xu2018temperature,mutch2019evidence}, and scenario 2 remains possible if one can reconcile the discrepancy between the temperature-independent $2|\Delta|$ at low temperatures with the theoretical prediction. Here, we propose a possible mechanism based on our recent discovery of a second band gap in the vicinity of $\Gamma$ point in the STI phase of ZrTe$_5$ \cite{Jiang2020unraveling,Wang2021NC}. In Figure \ref{fig4}c, we schematically depict the temperature evolution of the band structure across the TPT. At the elevated temperature above the TPT, the system is in the WTI phase, and the band structure takes the form of a conventional massive Dirac electron with only one extremum at $\Gamma$ point \cite{mutch2019evidence,Fan2017transition,Wang2020ultrafast}. The incident IR light thus probes the band gap $2|\Delta|$ at $\Gamma$ point. Once the temperature drops to the TPT temperature, the conduction and valence bands come close enough so that the orbital mixing effect starts to play a role and leads to an anticrossing gap of finite size \cite{Krizman2018avoided,Wojek2014band}. When the temperature is further reduced, the system is in the STI phase. Due to the competition between the band inversion and the linear dispersion component, the band structure develops a second band gap in the vicinity of $\Gamma$ point along the layer stacking direction ($b$ direction) \cite{Jiang2020unraveling,Wang2021NC,morice2020optical,Weng2014transition}. Such a band structure resembles that in a gapped Weyl node system, and the size of the second band gap is determined by the strength of hybridization between the conduction and valence bands in an inverted band structure \cite{Jiang2017probing}. Our previous calculation shows that the joint density of states across the second band gap could dominate that at $\Gamma$ point \cite{Jiang2020unraveling}. Therefore, the incident IR light no longer probes the band gap at $\Gamma$ point but the hybridization gap at its vicinity. In this situation, the temperature-independent $2|\Delta|$ observed in our experiment reflects the size of the hybridization gap, which remains a constant below the TPT. Our result does not violate the theoretical prediction about the increasing gap at $\Gamma$ point as lowering the temperature, and hence scenario 2 is still possible.

In fact, in our previous work \cite{Jiang2020unraveling}, samples grown by the same flux method are shown to be in the STI phase at low temperatures based on the observation of two sets of Dirac band LL transitions with similar magnetic field dependence. We attribute the additional set of transitions to those across the second band gap in the vicinity of $\Gamma$ point and consider it a unique indicator of the STI phase. Such a connection is consistent with the prior \textit{ab initio} calculations \cite{Weng2014transition,morice2020optical,zhu2022comprehensive} and has further been confirmed by a unique four-fold splitting in quantum oscillations \cite{Wang2021NC}, which excludes defect/impurities as a possible origin \cite{defects}. For the ZrTe$_5$ sample studied in this work, between the strong transitions labeled by LL index $n$ in Figure \ref{fig2}a,b, we also find a series of weak transitions indicated by the asterisk symbol ($*$) in close resemblance with our previous work \cite{Jiang2020unraveling}. Hence, these asterisk modes suggest the STI phase in ZrTe$_5$ at low temperatures. Consequently, a TPT possibly occurs between 30 K and 60 K as these modes start to lose their visibility and the band parameters start to fluctuate.

In conclusion, we have performed a magneto-IR spectroscopy study of the LL transitions in Te-flux grown ZrTe$_5$ samples in the temperature range of 6 K to 175 K and observed distinct temperature dependence for different low-energy modes. Using a massive Dirac fermion model, we extract the band gaps and Fermi velocities from the observed LL transitions and find that as temperature decreases, the band gap (Fermi velocity) first decreases (increases) and then saturates at low temperatures. We show that such a temperature dependence not only can be explained with a no TPT scenario but also is compatible with a TPT scenario if the formation of the second band gap (due to band inversion) and the associated orbital mixing effect are considered. We note that given the different topological phases reported for ZrTe$_5$ at low temperatures, both scenarios are not exclusive and may also be sensitive to sample details. Nevertheless, our work presents a new possibility for the band gap evolution across the TPT, which may be easily dismissed.

We thank David Singh for insightful discussions. This work was primarily supported by the DOE (Grant No. DE-FG02-07ER46451), while crystal growth at UTK was supported by Grant No. DE-SC0020254. Y. J. acknowledges the support by the National Natural Science Foundation of China (Grant No. 12274001) and the Nature Science Foundation of Anhui Province (Grant No. 2208085MA09). The crystal characterization was performed in part at the GT Institute for Electronics and Nanotechnology, a member of the National Nanotechnology Coordinated Infrastructure, which is supported by the NSF (Grant No. ECCS-1542174). The IR measurements were performed at NHMFL, which is supported by the NSF Cooperative Agreement No. DMR-1644779 and the State of Florida.

\textit{Note added.} After the initial submission of this work, we became aware of a similar work \cite{mohelsky2023temperature}, presenting consistent experimental results with the no-TPT interpretation.

\section{Appendix: Additional experimental data}
\begin{figure}[h!]
\centering
\includegraphics[width=3.35in] {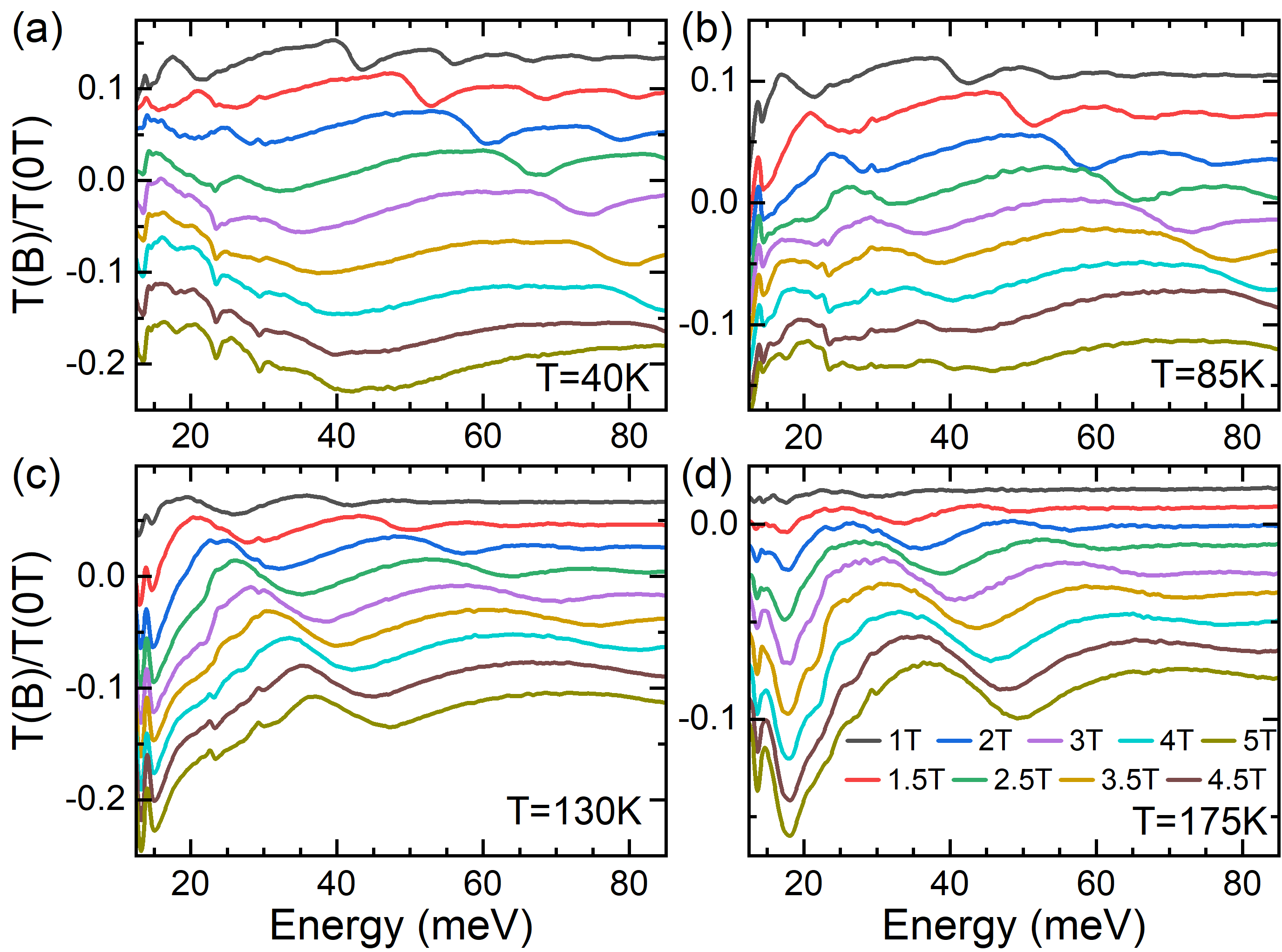}
\caption{Normalized magneto-transmission spectra measured from $B=1$ T to 5 T with a step size of 0.5 T under a fixed temperature. The selected temperatures are (a) 40 K, (b) 85 K, (c) 130 K, and (d) 175 K. In each panel, the spectra are offset vertically for clarity.}
\label{fig5}
\end{figure}

\begin{figure}[t!]
\centering
\includegraphics[width=3.35in] {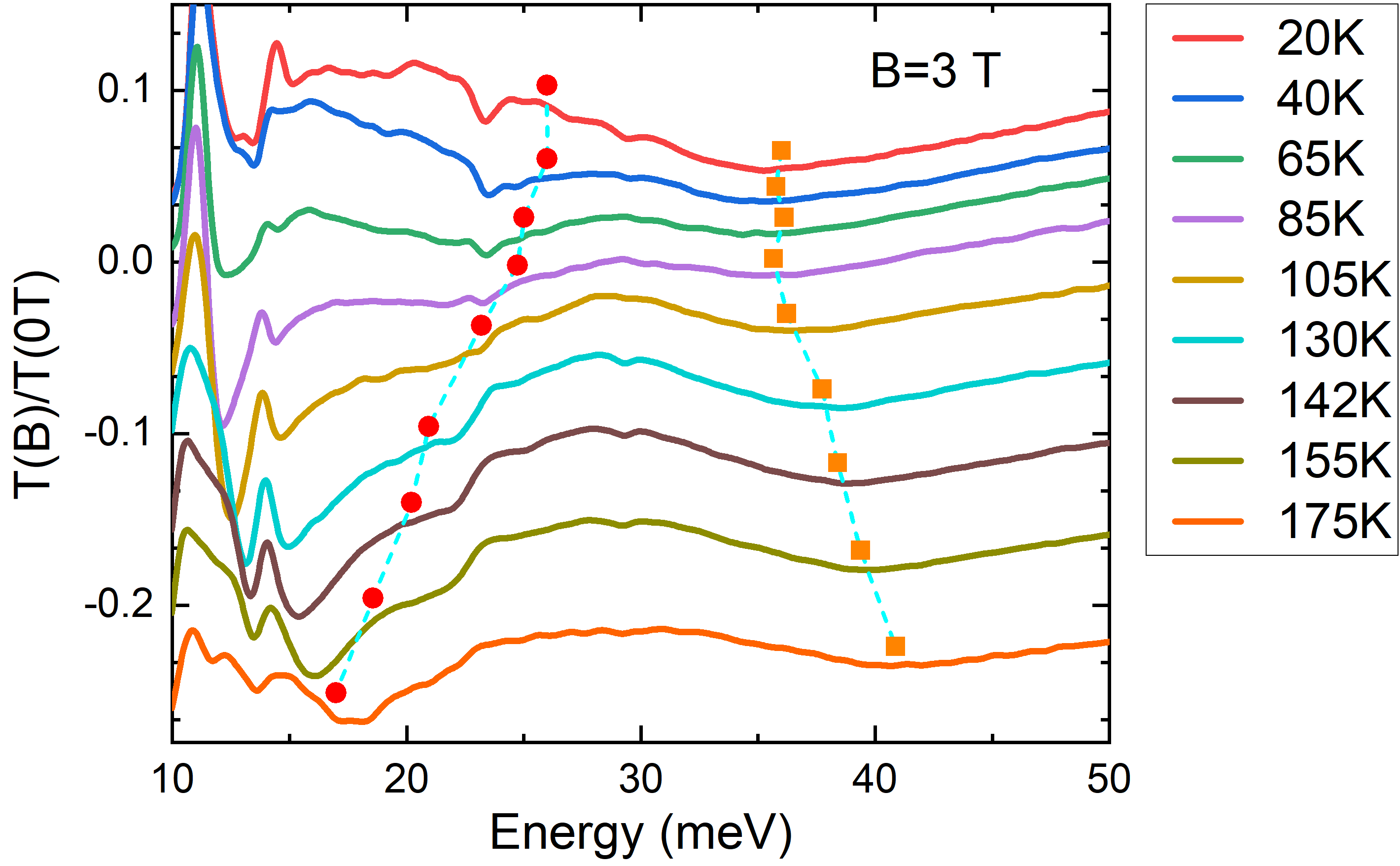}
\caption{Temperature-dependent normalized magneto-transmission spectra measured from 20 K to 175 K at 3 T. The spectra are offset vertically for clarity.}
\label{fig6}
\end{figure}

For completeness, in Figure \ref{fig5}, we include additional data describing the magnetic field evolution of the normalized magneto-transmission spectra at different temperatures. In Figure \ref{fig6}, we present the raw spectra of the false color plot of Figure \ref{fig2}c.

\bibliography{NL-sample}

\begin{thebibliography}{47}
\bibitem [{def()}]{defects}%
  \BibitemOpen
  \href@noop {} {}\bibinfo {note} {In addition, the energies of these weak
  modes do not fit well with other selection rules, such as $\Delta n=0,\pm 2$,
  induced by defects/impurities. Quantitative analysis of the optical weight
  and energy dependence of the weak modes can be found in Ref.
  [36].}\BibitemShut {Stop}%
\end{thebibliography}
\end{document}